\documentclass[english]{article}
\usepackage[T1]{fontenc}
\usepackage[utf8]{inputenc}
\usepackage{geometry}
\usepackage{xcolor}
\usepackage{array}
\usepackage{hyperref}
\usepackage{multirow}
\usepackage{graphicx}
\usepackage{float}
\usepackage{authblk}
\usepackage{amsmath}
\usepackage{subcaption}
\usepackage{soul}
\usepackage{textcmds}
\usepackage{graphicx}
\usepackage{textgreek}
\makeatletter

\begin{document}
\title{Remote Implementation of Hidden or Partially Unknown Quantum Operators using Optimal Resources: A Generalized View}
\author[1,$\#$]{Satish Kumar}
\author[1,$\dagger$]{Kuldeep Gangwar}
\author[1,$\ast$]{Anirban Pathak}
\affil[1]{Jaypee Institute of Information Technology, A 10, Sector 62, Noida, UP-201309, India}
\affil[$\ast$]{Corresponding author: anirban.pathak@gmail.com}
\affil[$\#$]{mr.satishseth@gmail.com}
\affil[$\dagger$]{kuldeep972018@gmail.com}

\maketitle
\begin{abstract}
 Two protocols are proposed for two closely linked but different variants of remote implementation of quantum operators of specific forms. The first protocol is designed for the remote implementation of the single qubit hidden quantum operator, whereas the second one is designed for the remote implementation of the partially unknown single qubit quantum operator. In both cases two-qubit maximally entangled state, which is entangled in the spatial degree of freedom is used. The quantum resources used here are optimal and easy to realize and maintain in comparison to the multi-partite or multi-mode entangled states used in earlier works. The impact of photon loss due to interaction with the environment is analyzed for both the schemes. The proposed protocols are also generalized to their controlled, bidirectional, cyclic, controlled cyclic, and controlled bidirectional versions and it is shown that either Bell state alone or products of Bell states will be sufficient to perform these tasks with some additional classical communications in the controlled cases only. This is in sharp contrast to the earlier proposals that require large entangled states.  In addition, it's noted that remote implementation of hidden or partially unknown operators involving multiple controllers and/or multiple players who jointly apply the desired operator(s) would require quantum channels more complex than the Bell states and their products. Explicit forms of such quantum channels are also provided.
\end{abstract}

\section{Introduction}

Entangled states can be used to realize tasks having no classical analog, like quantum teleportation of an unknown quantum state \cite{BB93}. Quantum teleportation was introduced in 1993, and since then several variants of it have been introduced \cite{Karlsson1998CQT,HVC+01,Chitra2013BCQT}. One of the most relevant variants is the remote state preparation or RSP that allows teleportation of a known quantum state. In particular cases, it requires a lesser amount of classical resources than that required for teleportation of an unknown quantum state \cite{P00}. Lately, RSP has also been extended to joint RSP \cite{BaAn2008JRSP}, controlled joint RSP \cite{SSB+15}, etc. Interestingly, the idea of RSP and its generalized versions can also be extended to a situation where instead of a state being prepared remotely, an operator is implemented remotely. Such a protocol is generally referred to as the protocol for the remote implementation of operator (RIO). The first scheme of RIO was proposed in 2001 \cite{HVC+01} and it clearly established that a quantum operation can be prepared remotely using shared entanglement along with local operation and classical communication (LOCC). Because of its analogy with conventional remote controls, RIO is also called quantum remote control. Interestingly, the ability to perform teleportation allows us to perform the task performed by RIO. Lucidly speaking Alice may teleport a state to Bob and Bob may apply an operator $U_B$ on the state received by him and teleport back the output state to Alice. Thus, a bidirectional quantum state teleportation  \cite{TVP15,SST+17} can be used to realize a scheme for RIO. Such a trivial realization would require a pair of shared Bell states and the communication of 4 classical bits. In contrast, it's possible to implement RIO with a lesser amount of resources if $U_B$ belongs to certain classes of unitary operators. Specifically, in \cite{HPV02}, it was also that in such a case (i.e., when $U_B$ belongs to certain classes of unitary operators),  RIO can be implemented using 1 Bell state and 2 bits of classical communication only. The concept of RIO is also experimentally realized for a specific version of RIO \cite{XLG05} and generalized in many ways. For example, we may mention joint RIO (JRIO) \cite{BAn22}, controlled RIO (CRIO) \cite{QC23,BB22} and controlled joint RIO (CJRIO) \cite{KP24}. All the tasks performed by these schemes can be realized using bidirectional quantum state teleportation and the controlled, joint, or controlled joint version of that, but there is a related task that cannot be realized in that way. It was identified by Ba An in 2007 with a task which is referred to as the remote implementation of a hidden operator (RIHO) \cite{A07}. The task will be described properly in Section \ref{sec:tasks}, but here we may note that Bob's operator $U_B$ can be decomposed as $U_B=U_0+U_1,$ where the suboperators $U_0$ and $U_1$ are of a specific form to be described in Section \ref{sec:tasks}. In RIHO, Alice possesses a quantum state $|\psi\rangle_X$ and needs the transformed state $U_0|\psi\rangle_X$ or $U_1|\psi\rangle_X$ but not $U_B|\psi\rangle_X$. The protocol allows Bob to remotely apply $U_B$ on $|\psi\rangle_X$, but at the end of the protocol, Alice randomly obtains either $U_0|\psi\rangle_X$ or $U_1|\psi\rangle_X$. 

 As we mentioned above, RIO is developed in analogy with RSP which is a variant of quantum teleportation where the state to be teleported is known. Thus, the basic difference between a conventional scheme for teleportation and that of RSP is that in the former case, the state to be teleported is unknown whereas in the latter case, it is known. Extending the analogy beyond an analog of RSP, can we think of a version of RIO which is closer to teleportation in the sense that the operator will be unknown? Specifically, it will be interesting to ask whether a scheme for remote implementation of partially known or unknown quantum operations (RIPUO) can be designed and whether such a scheme will have any advantage. Such a question was first introduced by Wang in 2006 \cite{W06}. In Ref. \cite{W06}, a scheme for RIPUO was introduced and it was shown that the scheme requires half of the resources compared to that used in bidirectional quantum state teleportation. The idea was further extended to the case of many senders and many controllers by Wang himself in 2007 \cite{W07}. Since then many variants of it have appeared, such as controlled remote implementation of partially unknown quantum operation (CRIPUO) \cite{FL08}, cyclic controlled remote implementation of partially unknown quantum operations (CCRIPUO) \cite{PH19}, bidirectional cyclic controlled remote implementation of partially unknown quantum operations (BCCRIPUO) \cite{PY+22} and many-party controlled remote implementation of partially unknown quantum operations (MCRIPUO) \cite{PT+22}. Different resources are used in realizing these protocols. For example, in Ref. \cite{A07} a tripartite entangled state was used to realize RIHO, and in Ref. \cite{W06} the Bell state was used to realize RIPUO. Different variants of RIHO (CRIHO \cite{A07}) and RIPUO (CRIPUO \cite{FL08}, CCRIPUO \cite{PH19}, BCCRIPUO \cite{PY+22}, MCRIPUO \cite{PT+22}) are realized with $3+n$ and $2+n$ party entangled state respectively, where $n$ are additional parties involved in comparison to the original schemes of RIHO \cite{A07} and RIPUO \cite{W06}. In what follows, we will show that not only RIHO, but even controlled RIHO (CRIHO), RIPUO, and CRIPUO can be realized using a bipartite entangled state. Specifically, keeping in mind, the problem of preparation and maintenance of larger entangled states, an effort has been made here to propose new protocols for a family of RIO tasks including RIHO, CRIHO, RIPUO, CRIPUO, etc. using the optimal amount of quantum resources\footnote{The notion of optimality used here is as follows. As entanglement is necessary for the realization of any version of the remote implementation of an operator, any scheme would require at least a bipartite or two-mode entanglement. Thus, any variant of RIO realized using a Bell state must be optimal. However, in more complex cases, we may need to check the minimum size of the required entangled states and number of copies of that state needed for the implementation of the protocol.}. We will also discuss: How to realize RIPUO and its variants using optimal quantum resources? Further, till now the scheme of RIHO and that of RIPUO are developed independently, but in what follows we will show that there is a common framework and some simple changes in the scheme of RIHO can transform it into a scheme for RIPUO.

The rest of this paper is organized as follows. In Section \ref{sec:tasks}, the problem statement is provided. Specifically, the general form of the operators that are to be implemented remotely (either partially or in a hidden manner) is described. Subsequently, in Section \ref{Sec:Protocols} two protocols for specific types of remote implementation of the operators are described. The first protocol is designed for the remote implementation of hidden quantum operators whereas the second protocol is proposed for the remote implementation of the partially unknown quantum operators. All these are done using the optimal amount of quantum resources only. In the same section, we also analyze photon loss of auxiliary coherent state due to coupling with the environment for both protocols. Proposed protocols are further generalized in Section \ref{sec:general-method}. Specifically, it's shown that the controlled version of the proposed schemes can be implemented without using any additional quantum resources. Further, it is shown that the proposed protocols can be generalized to their multiparty versions, where an arbitrary number of parties can jointly implement a protocol for the remote implementation of the desired hidden/partially unknown operators. The paper is concluded in Section \ref{sec:conclusion}.

\section{The tasks of our interest\label{sec:tasks}}
We have two communicating parties Alice and Bob in our proposed protocols, which are geographically far apart. Bob has an arbitrary unitary operator $U_B$ which can be visualized differently for different scenarios (tasks). In the first scenario, we may visualize the unitary $U_B$  as a "lump operator" as described in Ref. \cite{A07}. $U_B$ represents a unitary operator given by 
\begin{equation}\label{eq:U}
    U_B=\frac{1}{\sqrt{2}}\begin{pmatrix}
    u & v\\ 
    -v^{*} & u^{*}
    \end{pmatrix}
    \, = \frac{1}{\sqrt{2}}(\, U_0 \, + \, U_1),
\end{equation}
where
\begin{equation}\label{eq:U_i}
    U_0=\begin{pmatrix}
    u & 0\\ 
    0 & u^{*}
    \end{pmatrix}
    \quad \& \quad 
    U_1=\begin{pmatrix}
    0 & v\\ 
    -v^{*} & 0
    \end{pmatrix}.
\end{equation}

Here, the operator $U_0$ ($U_1$) commutes (anti-commutes) with Pauli-z operation $\sigma_z$, and the operators $U_0$ and $U_1$ can be described as diagonal and antidiagonal operators, respectively \cite{W06}. Note that if we consider $U_B$ to be unimodular (i.e., if we consider $|u|^2+|v|^2=2$), a set of all operators of the above form, i.e., ${U_B}$ would form $SU(2)$ group. Further, any rotations that can be implemented on a qubit can be described as a matrix of the above form. Thus, we may view the lump operator $U_B$ as a rotation which can be decomposed as $U_B=\frac{1}{\sqrt{2}}(U_0+U_1)$ where the sub-operators $U_0$ and $U_1$ can be viewed as restricted types of rotation and the imposition of the additional conditions $|u|=|v|=1$ would ensure the unitarity of $U_0$ and $U_1$ making it convenient to visualize them as physically realizable operators. To visualize this further, let us provide some simple examples. Let us recall that $R_{y}(\theta)$ and $R_{z}(\theta)$ denote single qubit quantum gates that rotate a qubit through an angle $\theta$ around the y-axis and z-axis, respectively. In the matrix form, these gates are written as 
\begin{equation}\label{eq:ry}
    R_{y}(\theta)=\begin{pmatrix}
    \cos(\frac{\theta}{2}) & -\sin(\frac{\theta}{2})\\ 
    \sin(\frac{\theta}{2}) & \cos(\frac{\theta}{2})
    \end{pmatrix}
    \quad \& \quad 
    R_{z}(\theta)=\begin{pmatrix}
    \exp{(-\iota\frac{\theta}{2})} & 0\\ 
    0 & \exp{(\iota\frac{\theta}{2})}
    \end{pmatrix}.
\end{equation}
Clearly, $R_{z}(\theta)$ and $R_{y}(\theta=0)$ are of the form of $U_0$, whereas $R_{y}(\theta=\pi)$ is of the form of $U_1$. These simple examples would help us to physically visualize the tasks that we wish to perform remotely. In fact, further elaborating on the above examples of the sub-operators $U_0$ and $U_1$, we may note that one can physically interpret the operator $U_0$ and $U_1$ as arbitrary rotations around the z-axis and rotations by $\pi$ around any axis in the equatorial plane, respectively \cite{HPV02}.

Now the first task is such that Bob can remotely apply $U_B$ on a single-qubit state $|\psi\rangle_X$ of Alice,  but he cannot apply $U_m$ $(m=0,1)$ separately. However, at the end of the scheme, Alice's state $|\psi\rangle_X$ gets transformed to either $U_0|\psi\rangle_X$ or $U_1|\psi\rangle_X$. This is called the remote implementation of the hidden operator as the operator which is actually implemented (i.e., the sub-operator $U_0$ or $U_1$) on the state of Alice, remains hidden inside the lump operator $U_B$ available with Bob.
 
In the second scenario (i.e., in the second task), the sub-operator $U_0$ or $U_1$ can be seen as a partially unknown operator as the values of the matrix elements are unknown to Bob, but the structure of the matrices or the position of non-zero matrix elements are known to him. Here, Bob applies $U_m$ $(m=0,1)$ on Alice's state $|\psi\rangle_X$ which transforms it to $U_m|\psi\rangle_X$ at the end of the scheme.\\
The arbitrary quantum state which Alice possesses can be described as follows:
\begin{equation}\label{eq:Qs}
    |\psi\rangle_X=(\alpha|x_0\rangle+\beta|x_1\rangle)_X,
\end{equation}
where $\alpha$ and $\beta$ are unknown coefficients which satisfy the normalization condition $|\alpha|^2+|\beta|^2=1$.\\
Depending on the unitary visualization for the two different scenarios, two different tasks of remote implementation of operators (RIO) have been described above. First, Bob applies unitary $U_m$ $(m=0,1)$ on Alice's unknown quantum state $|\psi\rangle_X$ when Bob has operator $U_B$. Second, Bob applies either $U_m$ $(m=0,1)$ on Alice's quantum state $|\psi\rangle_X$ when Bob has operator $U_m$ $(m=0,1)$. To perform both tasks, they share the two-qubit maximally entangled state, which is entangled in spatial degree of freedom (S-DOF) given as follows:
\begin{equation}\label{eq:QC}
    |\Omega^{+}\rangle_{AB}=|a_0\rangle_A |b_0\rangle_{B} + |a_1\rangle_A |b_1\rangle_{B}
\end{equation}
The first and second qubit in Eq. (\ref{eq:QC}) is with Alice and Bob, respectively. Note, that the normalization factor $1/\sqrt{2}$ is omitted in  Eq. (\ref{eq:QC}) and the subsequent equations for simplicity.\\
The steps involved in completing both tasks described here are explained in Section \ref{Sec:Protocols}. 
\section{Protocol for remote implementation of hidden or partially unknown quantum operators}\label{Sec:Protocols}
The joint state consists of Alice's unknown quantum state and the quantum channels used to complete both tasks can be written as follows:
\begin{equation}\label{eq:Joint}
    |\psi\rangle_X|\Omega^{+}\rangle_{AB}=(\alpha|x_0\rangle+\beta|x_1\rangle)_X 
\otimes(|a_0\rangle_A |b_0\rangle_{B} + |a_1\rangle_A |b_1\rangle_{B})
\end{equation}
The two cases of unitary which Bob has $(i)$ when Bob has access to $U_B$ but not to $U_m$ $(m=0,1)$ or $U_m$ is hidden to Bob and $(ii)$ when Bob knows only the structure of $U_m$ but not its value. For both cases, Bob is demanded to operate $U_m$ on Alice's unknown state $|\psi\rangle_X$. The protocols for case $(i)$ and case $(ii)$ are described in step-wise manner in Section \ref{sec:hidden} and Section \ref{sec:unknown}, respectively.

\subsection{Protocol for remote implementation of hidden quantum operators}\label{sec:hidden}   
\begin{figure}
    \centering
    \includegraphics[width=\textwidth]{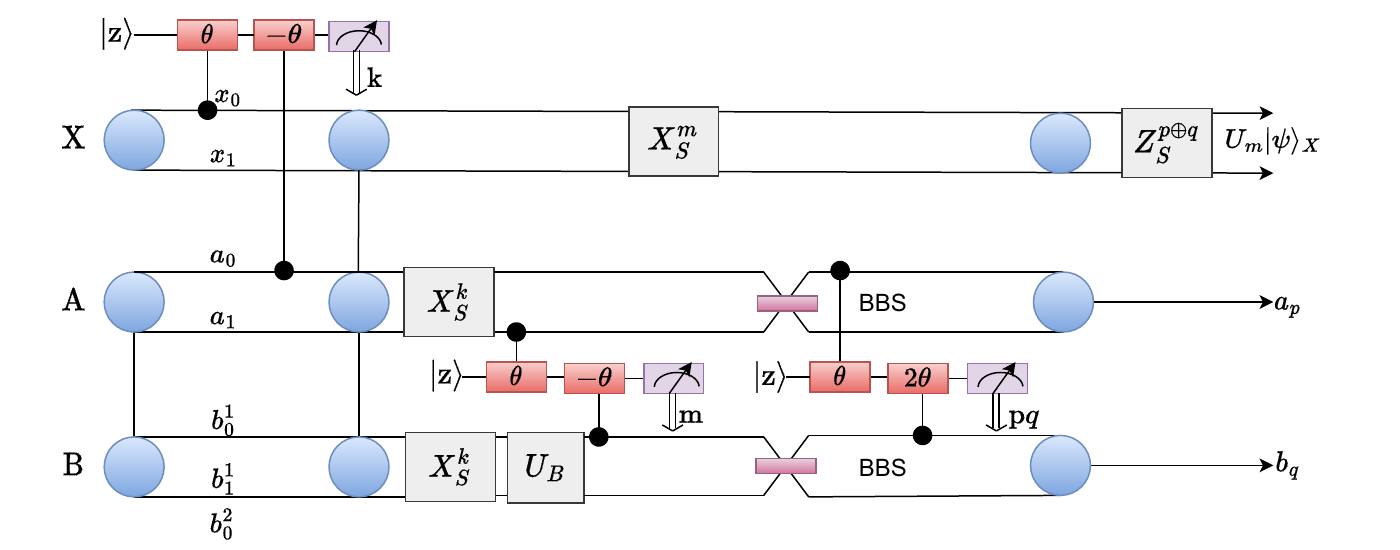}
    \caption{(Color Online) This figure describes the steps involved in the protocol for implementing hidden operators to remote parties. A circle represents the unpolarized photon. The parallel line attached to the circle represents two different paths for the photon. The solid line connecting the two circles represents the entanglement in S-DOF. The cross-Kerr interaction between a CS $|z\rangle$ and a photon path is represented by a solid line connecting them with a solid dot on the photon path. $\pm\theta$ and $2\theta$ are phase shifts in the CS. The double arrow coming out from the CS measurement box represents the measurement outcome. BBS here is a balanced beam splitter.} 
    \label{fig:Fig01}
\end{figure}

    Photonic setups provide us excellent platforms for the development of quantum technologies because photons are weakly interacting particles, and that leads to minimal interaction with the environment. This property is beneficial for maintaining long coherence times. However, a key requirement for quantum photonic technology is photon-photon interaction. Over the past years, significant efforts have been made to achieve strong interactions between two optical beams, both theoretically and experimentally \cite{FDS16,FHM+15,HWR+16,SKL+18,LCC+16,TSR+16,SSJ+20}. Interestingly, weak photon-photon scattering presents a significant challenge in achieving such interactions at the single-photon level.\\
    Despite this challenge, significant progress has recently been made in developing photonic quantum technologies including photonic quantum computers which require proper use of photon-photon interaction. For instance, cross-phase shifts of $13$ $\mu$rad per photon have been achieved through the implementation of a strong cross-Kerr non-linearity using electromagnetically induced transparency (EIT) \cite{FHM+15}. Additionally, cross-phase shifts of $0.3$ mrad per photon have been reported using rubidium atoms confined to a hollow-core photonic bandgap fiber \cite{VSG+13}, and shifts of $3$ mrad per polariton have been observed with exciton polaritons in micropillars  with embedded quantum wells. Beyond cross-Kerr interactions, a phase shift of $\pi$ between photons has been reported through interactions with atom-cavity systems \cite{HWR+16,SKL+18} and EIT \cite{TSR+16,LCC+16,SSJ+20}.\\
    Keeping the above information in the background, in what follows, we will first briefly discuss how the auxiliary coherent state (CS) $|z\rangle$ acquires phase shift $\theta$ via cross-Kerr interaction. In order to achieve this, we use cross-Kerr interaction between signal mode and probe mode. The interaction Hamiltonian between two modes $a$ and $b$ is $H=\hbar\chi a^{\dagger} a b^{\dagger} b$, where $\chi$ is non-linear coupling constant and $a(b)$ and $a^{\dagger}(b^{\dagger})$ are annihilation and creation operators of the mode $a$ and mode $b$, respectively. We may now consider the states of signal mode $a$ and probe mode $b$ are Fock state $|n\rangle$ and coherent state $|z\rangle$, respectively. Thus, the joint state of the two modes would evolve into $e^{-i \chi a^{\dagger} a b^\dagger b t}|n\rangle |z\rangle=|n\rangle|ze^{-i n \theta}\rangle$, where $\theta=\chi t$, and $t$ is the interaction time required to achieve the desired phase. We can notice that the signal mode phase is unaffected, while the probe mode acquires a phase shift $\theta$. Physical insights provided in this section till now may now be used to describe the desired protocol in a step-wise manner with appropriate attention to the physical process involved.
\begin{description}
    \item[Step1] First Alice entangles her qubit of unknown state $|\psi\rangle_X$ with the quantum channel $|\Omega^{+}\rangle_{AB}$. In order to entangle the qubit, she allows cross-Kerr interaction between CS $|z\rangle$ and the path $x_0$ of her photon X first, and then between the CS $|z\rangle$ and path $a_0$ of her photon A. The joint state Eq. (\ref{eq:Joint}) after first cross-Kerr interaction between $x_0$ and $|z\rangle$ with phase shift $\theta$ will evolve into $\alpha|x_0\rangle |a_0\rangle_A |b_0\rangle_{B}|z e^{i\theta}\rangle  +\beta|x_1\rangle |a_0\rangle_A |b_0\rangle_{B}|z\rangle+
   \alpha|x_0\rangle |a_1\rangle_A |b_1\rangle_{B} |z e^{i\theta}\rangle+\beta|x_1\rangle |a_1\rangle_A |b_1\rangle_{B} |z\rangle$.  The joint state Eq. (\ref{eq:Joint}) after second cross-Kerr interaction between $a_0$ and $|z\rangle$ with phase shift $-\theta$ will evolve into $(\alpha|x_0\rangle_X |a_0\rangle_A |b_0\rangle_{B} + \beta|x_1\rangle_X |a_1\rangle_A |b_1\rangle_{B})|z\rangle+ \alpha|x_0\rangle_X |a_1\rangle_A |b_1\rangle_{B}|ze^{i\theta}\rangle + \beta|x_1\rangle_X |a_0\rangle_A |b_0\rangle_{B}|ze^{-i\theta}\rangle$. To get the desired entangled state, she measures $\hat{\mathcal{X}}=a^\dagger+a$ quadrature of the CS $|z\rangle$ using homodyne measurement. We can consider $z$ as real without loss generality. The $\hat{\mathcal{X}}$ quadrature measurement is achieved by projecting the CS $|z\rangle$ into $|\mathcal{X}\rangle\langle\mathcal{X}|$
\begin{equation}
     	\begin{split}
        |\xi\rangle= & f(x,z)(\alpha|x_0\rangle_X |a_0\rangle_A |b_0\rangle_{B} + \beta|x_1\rangle_X |a_1\rangle_A |b_1\rangle_{B})\\
        	& + f(x,z\cos{\theta})(\alpha|x_0\rangle_X |a_1\rangle_A |b_1\rangle_{B}e^{i \phi(x)} + \beta|x_1\rangle_X |a_0\rangle_A |b_0\rangle_{B}e^{-i \phi(x)}),
        \end{split}
    \end{equation}
where $f(x,z)=(1/2\pi)^{1/4}\exp[{-\frac{1}{4}(x-2z)^2}]$, $f(x,z\cos{\theta})=(1/2\pi)^{1/4}\exp[{-\frac{1}{4}(x-2z\cos{\theta})^2}]$ and $\phi(x)=(x-2z\cos{\theta})z\sin{\theta}$. It is to be noted that $|ze^{ i\theta}\rangle$ and $|ze^{- i\theta}\rangle$ have same Gaussian curves $f(x,z\cos{\theta})$, therefore they cannot be distinguished by homodyne measurement. However, $|z\rangle$ and $|z e^{\pm i \theta}\rangle$ have two different Gaussian curves, and they can be distinguished as long as the separation between the peaks is large enough $(\mathcal{X}_d\sim z\theta^2\gg 1)$. The separation and the mid point between the peaks are $\mathcal{X}_d=2z(1-\cos{\theta})$ and $\mathcal{X}_0=z(1+\cos{\theta})$, respectively. If the  measurement outcome $\mathcal{X}$ is such that $\mathcal{X}>\mathcal{X}_0$, she obtains
\begin{align}
   |\xi^{'} \rangle= \alpha|x_0\rangle_X |a_0\rangle_A |b_0\rangle_{B} + \beta|x_1\rangle_X |a_1\rangle_A |b_1\rangle_{B}.
   \label{eq:8}
\end{align} 
On the other hand, if the measurement outcome is $\mathcal{X}<\mathcal{X}_0$, she gets
\begin{align}
    |\xi^{''} \rangle=\alpha|x_0\rangle_X |a_1\rangle_A |b_1\rangle_{B}e^{i \phi(x)} + \beta|x_1\rangle_X |a_0\rangle_A |b_0\rangle_{B}e^{-i \phi(x)}.
    \label{eq:9}
\end{align}
As some part of both the Gaussian curves $f(x,z)$ and $f(x,z\cos{\theta})$ are overlapping,  misidentification of the states $|\xi^{'} \rangle$ and $|\xi^{''} \rangle$ may happen. Probability of the happening of misidentification of the state may be referred to as the  error probability. This error probability is given as
\begin{align}
 P_{\rm{error}}= \frac{1}{2} \text{erfc}[\mathcal{X}_d/2\sqrt{2}],
 \label{eq:10}
\end{align}
where $\text{erfc}$ is a complimentry error function. In this present case, the error probability is $ P^1_{\rm{error}}= \frac{1}{2} \text{erfc}[z(1-\cos{\theta})/\sqrt{2}]$.  If she gets state $|\xi^{''} \rangle$, both Alice and Bob apply classical feed-forward operation \cite{MN05,NM2004cross-kerr} to eliminate phase shift $\phi(x)$ and the path-flip operator $X_S$ $(X_S=|a_0\rangle\langle a_1|+|a_1\rangle\langle a_0|)$ on their photons $\text{A}$ and $\text{B}$, otherwise (i.e., if $|\xi^{'} \rangle$ is obtained) they do not need to do anything. The joint state after the measurement becomes entangled which is given as follows:
\begin{equation}
    |\xi^{'''}\rangle=\alpha|x_0\rangle_X |a_0\rangle_A |b_0\rangle_{B} + \beta|x_1\rangle_X |a_1\rangle_A |b_1\rangle_{B}
\label{eq:11}
\end{equation}
\item[Step2] Now Bob applies an operator $U_B$ on his qubit $\text{B}$. After application of this operator, the state in Eq.\eqref{eq:11} transforms into the following:
\begin{equation}
    |\xi^{''''}\rangle =(I\otimes I\otimes U_B)|\xi^{'''}\rangle=\frac{1}{\sqrt{2}}[\alpha |x_0\rangle_X |a_0\rangle_A (u|b_0\rangle-v^{*}|b_1\rangle)_{B} + \beta|x_1\rangle_X |a_1\rangle_A (v|b_0\rangle+u^{*}|b_1\rangle)_{B}].
    \label{eq:12}
\end{equation}
where $I$ is the $2\times 2$ identity matrix.
\item[Step3] Alice takes an auxiliary CS $|z\rangle$, and allows it to interact with path $a_0$ of her photon $\text{A}$ via cross-Kerr interaction with phase shift $\theta$, the state described by Eq.\eqref{eq:12} evolves into $\frac{1}{\sqrt{2}}[\alpha |x_0\rangle_X |a_0\rangle_A (u|b_0\rangle-v^{*}|b_1\rangle)_{B}|z e^{i\theta}\rangle + \beta|x_1\rangle_X |a_1\rangle_A (v|b_0\rangle+u^{*}|b_1\rangle)_{B}|z\rangle]$. And, then Bob uses the same auxiliary CS $|z\rangle$ and lets the CS interact with path $b_0$ of his photon $\text{B}$ via cross-Kerr interaction with phase shift $-\theta$. After this second cross-kerr interaction, the state described by Eq.\eqref{eq:12} evolves into
\begin{align}
    |\xi^{'''''}\rangle=&\frac{1}{\sqrt{2}}[(\alpha u|x_0\rangle_X |a_0\rangle_A |b_0\rangle_{B} + \beta u^{*}|x_1\rangle_X |a_1\rangle_A |b_1\rangle_{B})|z\rangle+(-\alpha v^{*}|x_0\rangle_X |a_0\rangle_A |b_1\rangle_{B}|ze^{ i\theta}\rangle \nonumber \\
    &+ \beta v |x_1\rangle_X |a_1\rangle_A |b_0\rangle_{B}|ze^{- i\theta}\rangle)]
    \label{eq:13}
\end{align}
Subsequently, Bob performs $\hat{\mathcal{X}}$ quadrature measurement on CS. As we have already discussed that $|ze^{\pm i\theta}\rangle$ cannot be distinguished, but $|z\rangle$ and $|ze^{\pm i\theta}\rangle$ can be distinguished in measurement with error probability $ P^2_{\rm{error}}= \frac{1}{2} \text{erfc}[z(1-\cos{\theta})/\sqrt{2}]$. If he gets $(m=0)$ $|z\rangle$, he does nothing. On the other hand, if he gets $(m=1)$ $|ze^{\pm i\theta}\rangle$, he applies classical feed-forward to remove phase shift $\pm\phi(x)$ and the path flip operator on her photon $\text{X}$. After these operations, two states corresponding to two different outcomes are 
\begin{equation}
     |\xi_m\rangle = \frac{1}{\sqrt{2}}\left\{\begin{matrix}
\alpha u|x_0\rangle_X |a_0\rangle_A |b_0\rangle_{B} + \beta u^{*}|x_1\rangle_X |a_1\rangle_A |b_1\rangle_{B} & \text{if}\, m=0\\ 
-\alpha v^{*}|x_1\rangle_X |a_0\rangle_A |b_1\rangle_{B} + \beta v |x_0\rangle_X |a_1\rangle_A |b_0\rangle_{B} & \text{if}\, m=1
\end{matrix}\right.
\label{eq:14}
\end{equation}
\item[Step4] Alice and Bob mix paths of their photons $\text{A}$ and $\text{B}$, respectively using a balanced beam splitter (BBS). The BBS transformation rule is $|a_i\rangle=\frac{1}{\sqrt{2}}[|a_i\rangle+(-1)^i|a_{i\oplus1}\rangle]$ $(i=0,1)$ (for more details see \cite{TP19beamSplitter, BB22}). The mixing of path $a_0$ and $a_1$ ($b_0$ and $b_1$) of photon $\text{A}$ ($\text{B}$) results into
\begin{equation}
 |\xi^{\prime}_m\rangle=\frac{1}{2}\left\{\begin{matrix}(|a_0\rangle|b_0\rangle+|a_1\rangle|b_1\rangle)(\alpha u|x_0\rangle+\beta u^{*}|x_1\rangle)+(|a_0\rangle|b_1\rangle+|a_1\rangle|b_0\rangle)(\alpha u|x_0\rangle-\beta u^{*}|x_1\rangle) & \text{if}\, m=0\\(|a_0\rangle|b_0\rangle+|a_1\rangle|b_1\rangle)(-\alpha v^{*}|x_1\rangle+\beta v |x_0\rangle)+((|a_0\rangle|b_1\rangle+|a_1\rangle|b_0\rangle)(-\alpha v^{*}|x_1\rangle-\beta v |x_0\rangle) & \text{if}\, m=1
 \end{matrix}\right.
\label{eq:15}
\end{equation}
She then turns on the cross-Kerr interaction between an auxiliary CS $|z\rangle$ and path $|a_1\rangle$ with phase shift $\theta$ and forwards it to Bob, who also turns on the interaction between the CS and path $|b_1\rangle$ with phase shift $2\theta$, and performs $\hat{\mathcal{X}}$ measurement on it. The measurement outcomes are $pq=00,01,10,11$ corresponding to $|z\rangle,|ze^{i\theta}\rangle,|ze^{i2\theta}\rangle,|ze^{i3\theta}\rangle$ respectively. As adjacent Gaussian curves are overlapping corresponding to $|z\rangle,|ze^{i\theta}\rangle,|ze^{i2\theta}\rangle,|ze^{i3\theta}\rangle$, this may lead to misidentification. The error probabilities are $ P^{31}_{\rm{error}}= \frac{1}{2} \text{erfc}[z(1-\cos{\theta})/\sqrt{2}]$, $ P^{32}_{\rm{error}}= \frac{1}{2} \text{erfc}[z(\cos{\theta}-\cos{2\theta})/\sqrt{2}]$ and $ P^{33}_{\rm{error}}= \frac{1}{2} \text{erfc}[z(\cos{2\theta}-\cos{3\theta})/\sqrt{2}]$ of misidentification of $|z\rangle$ from $|ze^{i\theta}\rangle$, $|ze^{i\theta}\rangle$ from $|ze^{i2\theta}\rangle$ and $|ze^{i2\theta}\rangle$ from $|ze^{i3\theta}\rangle$ respectively. If $pq=01$ or $10$, then Alice applies the phase flip operator $(Z_S=|x_0\rangle\langle x_0|-|x_1\rangle\langle x_1|)$ on her photon $\text{X}$, otherwise (i.e., if $pq=00$ or $11$) nothing should be done. Finally, Bob has successfully been able to implement the unitary $U_m$ on Alice's arbitrary quantum state $|\psi\rangle_X$ and the normalized combined state is obtained as follows
\begin{equation}\label{eq:16}
        |\varphi\rangle=|a_p\rangle_A |b_q\rangle_{B}(U_m |\psi\rangle_X)
\end{equation}
with success probability
\begin{align}
    P_{\rm{1Suc}}=1- P_{\rm{error}}^1P_{\rm{error}}^2(P_{\rm{error}}^{31}+P_{\rm{error}}^{32}+P_{\rm{error}}^{33}).
    \label{eq:17}
\end{align}
\end{description}    
It is to be noted that Bob has applied the operator $U_B$ but Alice got $U_m|\psi\rangle_X$ $(m=0,1)$. Since the outcome $m$ is publicly announced Alice is firmly aware of which state, $U_0|\psi\rangle_X$ or $U_1|\psi\rangle_X$, she got. So, the task of implementing hidden operators becomes successful. The various steps involved in completing the task are pictorially represented in Fig. \ref{fig:Fig01}. One can also perform the task of RIHO using different quantum channels which are orthogonal to the channel used in Eq. (\ref{eq:QC}). The changes in the steps involved while using different quantum channels are shown in Table \ref{tab:stepsRIHO}. In all these cases, essentially a Bell state is used as a quantum channel. The use of a Bell state is optimal as entanglement is the required quantum resource for any RIO task and without maximal entanglement deterministic protocols would not work with unit fidelity. The point is that the ability to perform the desired tasks with different Bell states with a small change in the specific step(s), ensures that a third party (Charlie) may control the protocols of our interest just by randomly preparing Bell states and not distributing that among Alice and Bob without disclosing which state is shared till a later step. This point will be elaborated in Section \ref{sec:general-method}. However, before we do that we will describe a protocol for RIPUO. 
\begin{table}
\centering
\begin{tabular}{|c|c|c|c|}
\hline 
Steps & $|\Omega^{-}\rangle$ & $|\Pi^{+}\rangle$ & $|\Pi^{-}\rangle$\\
\hline 
Step1 & Same & $X_S^{k\oplus1}$ & $X_S^{k\oplus1}$\\
\hline
Step2 & Same & Same & Same\\
\hline
Step3 & Same & Same & Same\\
\hline
Step4 & $Z^{p\oplus q\oplus1}_S$ & Same & $Z^{p\oplus q\oplus1}_S$\\
\hline 
\end{tabular}
\caption{Variation in operations at several steps when using different quantum channels for the proposed protocol for RIHO. Here, $|\Omega^{\pm}\rangle=|a_0\rangle|b_0\rangle\pm|a_1\rangle|b_1\rangle$, $|\Pi^{\pm}\rangle=|a_0\rangle|b_1\rangle\pm|a_1\rangle|b_0\rangle$.}
\label{tab:stepsRIHO}
\end{table}

\subsection{Protocol for remote implementation of partially unknown quantum operators}\label{sec:unknown}
\begin{figure}
    \centering
    \includegraphics[width=\textwidth]{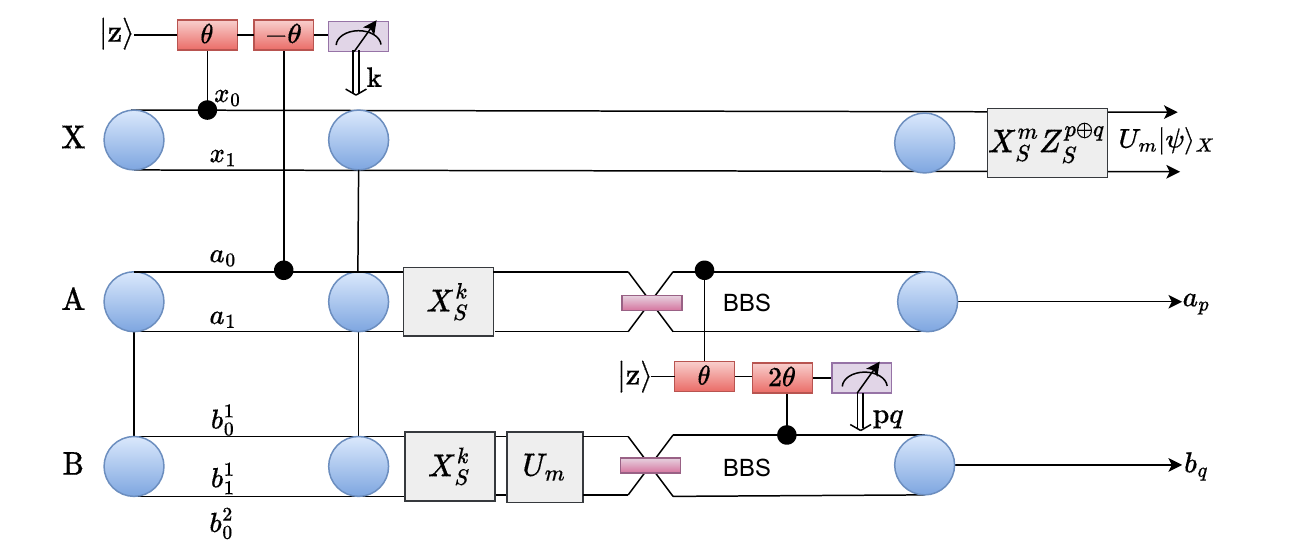}
    \caption{(Color Online) This figure describes the steps involved in the protocol for implementing a partially unknown operator to a remote party.}
    \label{fig:Fig02}
\end{figure}

\begin{description}
    \item[Step1] The first step is the same as Step1 of Section \ref{sec:hidden}.
    \item[Step2] Bob can either apply $U_0$ or $U_1$ on his photon state depending on his choice. If Bob applies $U_0$ then Eq. (\eqref{eq:11}) transforms to the following:
     \begin{equation}\label{eq:5}
        |\zeta_{0}\rangle=(I\otimes I\otimes U_0)|\xi^{'''}\rangle=\alpha u|x_0\rangle_X |a_0\rangle_A |b_0\rangle_{B} + \beta u^{*}|x_1\rangle_X |a_1\rangle_A |b_1\rangle_{B}
    \end{equation}
    and if Bob applies $U_1$ then Eq. (\eqref{eq:11}) transforms as follows.
    \begin{equation}\label{eq:6}
        |\zeta_{1}\rangle=(I\otimes I\otimes U_1)|\xi^{'''}\rangle=-\alpha v^{*}|x_0\rangle_X |a_0\rangle_A |b_1\rangle_{B} + \beta v|x_1\rangle_X |a_1\rangle_A |b_0\rangle_{B}
    \end{equation}
    \item[Step3] This step is same as Step4 of Section \ref{sec:hidden} with slight difference after Bob's measurement on the CS. If the measurement outcomes are $pq$ then Alice applies $X_{S}^{m}Z^{p\oplus q}_{S}$, where $m=0,1$ is the choice of unitary $U_m$ which Bob applied in the previous step. Finally, Bob is able to implement $U_m$ on Alice's state $|\psi\rangle_X$ and thus to yield     
    \begin{equation}\label{eq:7}
        |\varsigma\rangle=|a_p\rangle_A |b_q\rangle_{B}(U_m |\psi\rangle_X) 
    \end{equation}
    with success probability 
    \begin{align}
    P_{\rm{2Suc}}=1- P_{\rm{error}}^1(P_{\rm{error}}^{31}+P_{\rm{error}}^{32}+P_{\rm{error}}^{33}),
    \label{eq:21}
    \end{align}
    where $P_{\rm{error}}^1$ and $P_{\rm{error}}^{3i}$, $i=1,2,3$, are the same probabilities occurring in the Step1 and Step4 of the RIHO protocol, respectively.
\end{description}
One can see that Bob has applied the operator $U_m$ $(m=0,1)$ and Alice got the $U_m|\psi\rangle_X$ after following certain steps of the proposed protocol in Section \ref{sec:unknown}. So, the task of implementing partially unknown operators becomes successful. The various steps involved in completing the task are pictorial represented in Fig. \ref{fig:Fig02}. As was the case of RIHO, one can also perform the task of RIPUO using different quantum channels which are orthogonal to the channel described in Eq. (\ref{eq:QC}). The changes in the steps while using different quantum channels are shown in Table \ref{tab:stepsRIPUO}.
\begin{table}
\centering
\begin{tabular}{|c|c|c|c|}
\hline 
Steps & $|\Omega^{-}\rangle$ & $|\Pi^{+}\rangle$ & $|\Pi^{-}\rangle$\\
\hline 
Step1 & Same & $X_S^{k\oplus1}$ & $X_S^{k\oplus1}$\\
\hline
Step2 & Same & Same & Same\\
\hline
Step3 & $X_{S}^{m}Z^{p\oplus q\oplus1}_S$ & Same & $X_{S}^{m}Z^{p\oplus q\oplus1}_S$\\

\hline 
\end{tabular}
\caption{Variation in operations at several steps when using different quantum channels for the proposed protocol for RIPUO. Here, $|\Omega^{\pm}\rangle=|a_0\rangle|b_0\rangle\pm|a_1\rangle|b_1\rangle$, $|\Pi^{\pm}\rangle=|a_0\rangle|b_1\rangle\pm|a_1\rangle|b_0\rangle$.}
\label{tab:stepsRIPUO}
\end{table}
\subsection{Dissipation of auxiliary coherent state}
\begin{figure}
\centering
\begin{subfigure}{0.5\textwidth}
  \centering
  \includegraphics[width=0.8\linewidth]{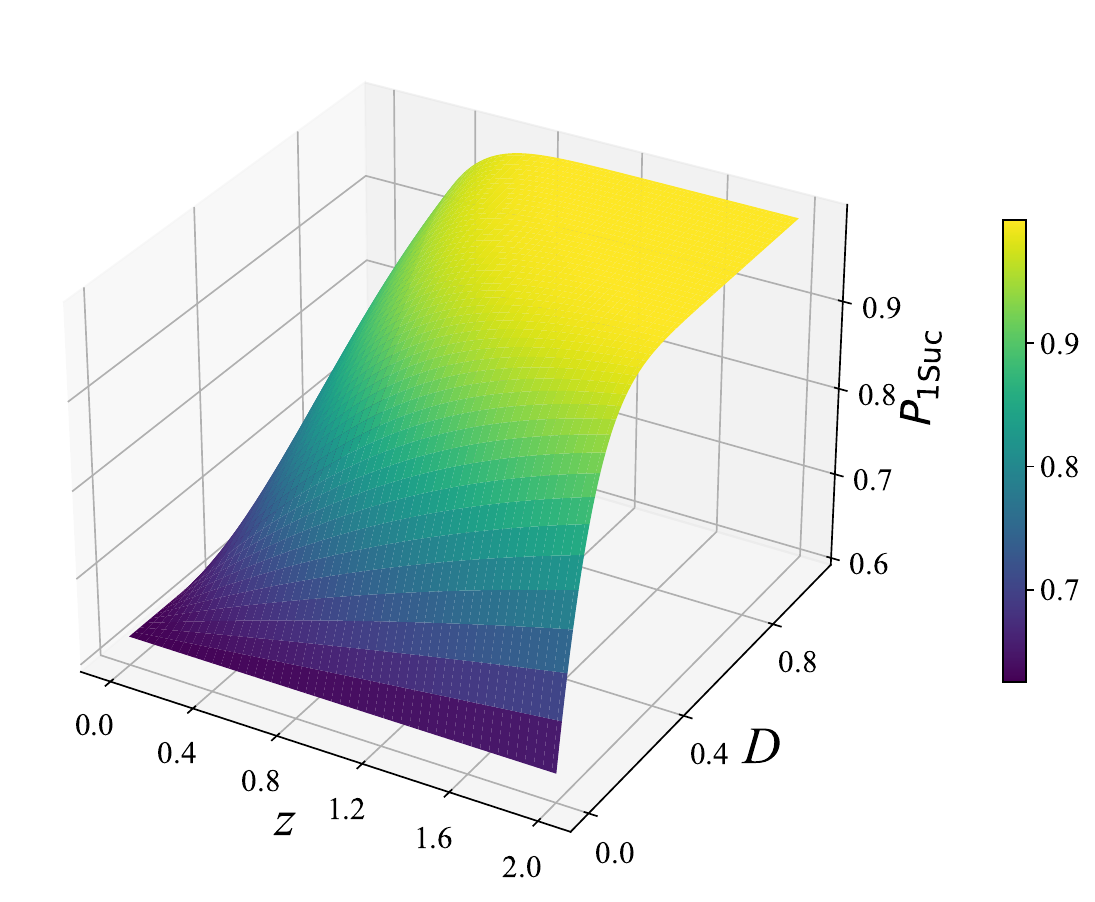}
  \caption{}
  \label{fig:sub-fig-a}
\end{subfigure}%
\begin{subfigure}{0.5\textwidth}
  \centering
  \includegraphics[width=0.8\linewidth]{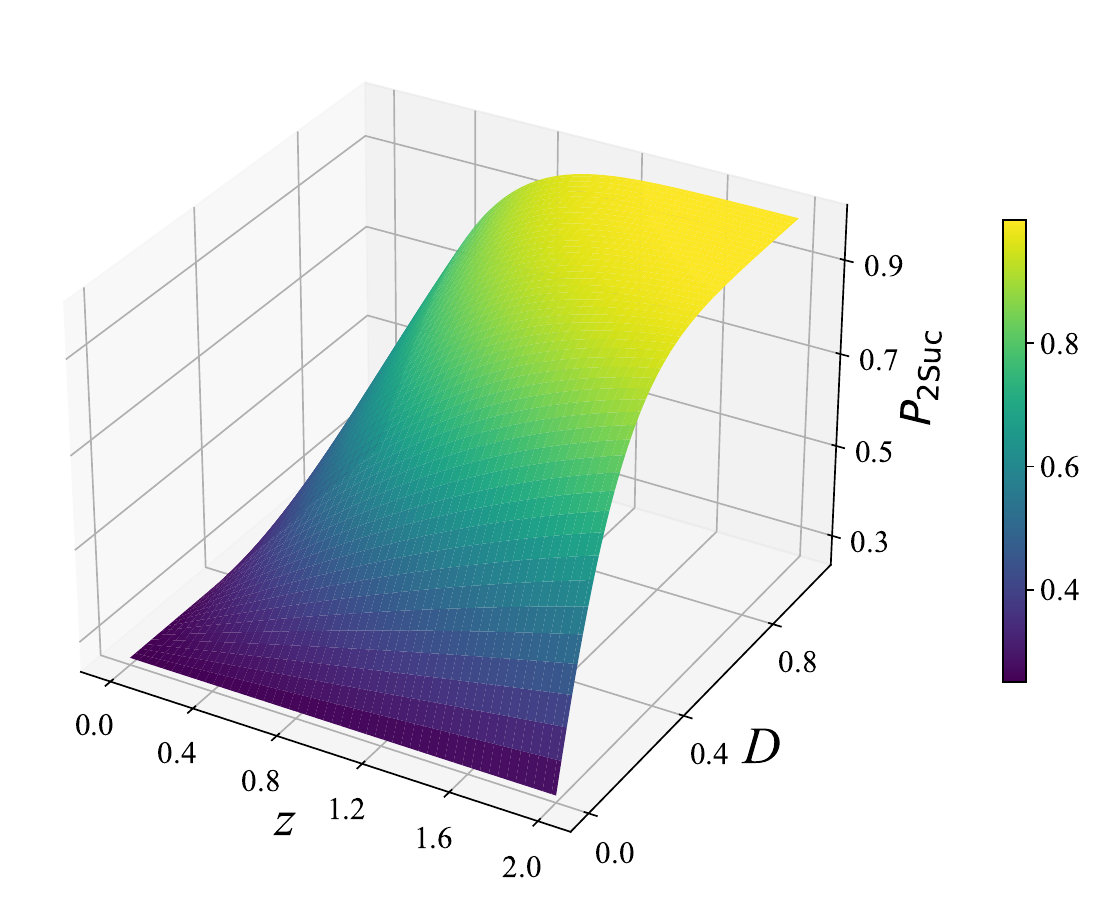}
  \caption{}
  \label{fig:sub-fig-b}
\end{subfigure}
\\[1ex] 
\begin{subfigure}{.5\textwidth}
  \centering
  \includegraphics[width=0.8\linewidth]{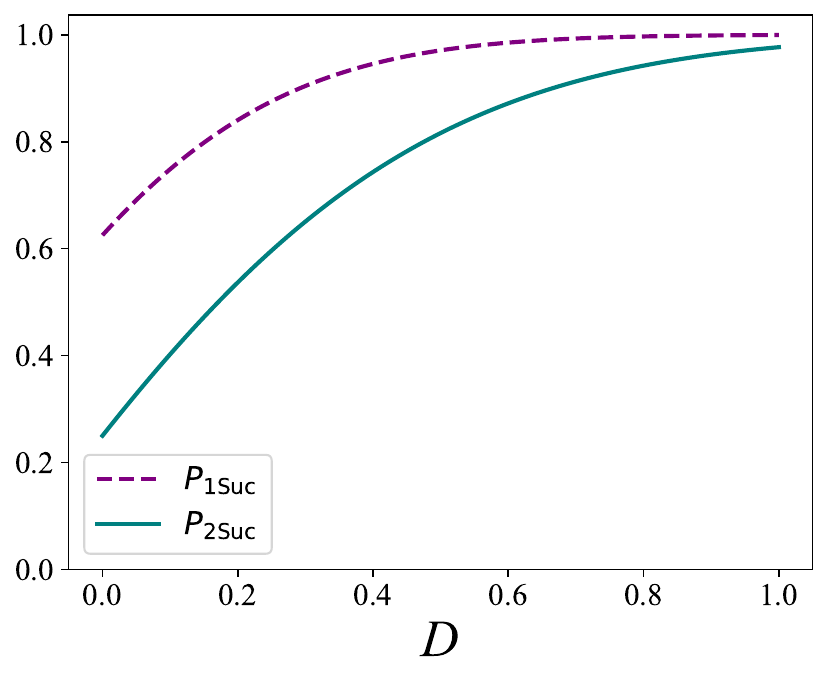}
  \caption{}
  \label{fig:sub-fig-c}
\end{subfigure}
\caption{(Color online) In (a) and (b) The behavior of modified success probabilities of RIHO ($P_{\rm{1Suc}}$) and RIPUO ($P_{\rm{2Suc}}$) protocols are shown as functions of dissipative parameter $D$  and initial amplitude $z$ of coherent state, with phase shift $\theta=\pi$ rad. In (c), the behavior of the same $P_{\rm{1Suc}}$ and $P_{\rm{2Suc}}$ with $D$ is shown, with phase shift $\theta=\pi$ rad and initial amplitude of the coherent state $z=1$.}
\label{fig:Fig03}
\end{figure}

In experimental setup, CS $|z\rangle$ is coupled with the environment. This undesired coupling with the environment may lead to the loss of some photons to the environment. Therefore, it is important to study the effect of dissipative coherent state $|z\rangle$ on $\hat{\mathcal{X}}$ quadrature measurement. When a quantum system interacts with its environment, the dynamics can be described by a master equation. In the interaction picture, the master equation provides a way to account for the system's evolution due to its coupling with the environment, focusing on how this interaction affects the system's density matrix $\rho(t)$. The master equation in interaction picture \cite{SB,BP07,GZ04,l20} is
\begin{align}
    \partial\rho(t)=\frac{\gamma}{2}[2b\rho(t)b^\dagger-\rho(t)b^\dagger b-b^\dagger b \rho(t)],
    \label{eq:28}
\end{align}
where $\rho(t)$ stands for density matrix of CS $|z\rangle$ at any time $t$, $b (b^\dagger)$ is anihilation (creation) operator of coherent state and $\gamma$ is dissipation constant for coherent state. \\
We can solve Eq. (\eqref{eq:28}) using the method described in the References \cite{P90, BK, J06}. All the Gaussian curves are modified to form $f(x, Dz\cos{(n\theta)})=(1/2\pi)^{1/4}\exp[{-\frac{1}{4}(x-2Dz\cos{(n\theta)})^2}]$, where $D=e^{-\gamma t}$ and $n=0,\pm1,\pm2,\pm3$. This modification in the Gaussian curves also alters the error probabilities. In fact, the error probabilities are modified to   $P^{1d}_{\rm{error}}= \frac{1}{2} \text{erfc}[Dz(1-\cos{\theta})/\sqrt{2}]$, $P^{2d}_{\rm{error}}= \frac{1}{2} \text{erfc}[Dz(1-\cos{\theta})/\sqrt{2}]$, $P^{31d}_{\rm{error}}= \frac{1}{2} \text{erfc}[Dz(1-\cos{\theta})/\sqrt{2}]$, $P^{32d}_{\rm{error}}= \frac{1}{2} \text{erfc}[Dz(\cos{\theta}-\cos{2\theta})/\sqrt{2}]$ and $P^{33d}_{\rm{error}}= \frac{1}{2} \text{erfc}[Dz(\cos{2\theta}-\cos{3\theta})/\sqrt{2}]$. As a consequence, success probabilities, Eq.\eqref{eq:17} and Eq.\eqref{eq:21}, are modified. Now, to visualize the impact of the dissipation process on the success probabilities, we may plot the variation of success probabilities with different physical parameters. The same is done in Fig. \hyperref[fig:Fig03]{3}

In Fig. \hyperref[fig:Fig03]{3(a,b)}, we plot the modified success probabilities of RIHO $P_{\rm{1Suc}}$ and RIPUO $P_{\rm{2Suc}}$ with dissipative parameter $D=e^{-\gamma t}$ and initial amplitude $z$ of coherent state $|z\rangle$, with phase shift $\theta$. We can observe that the initial amplitude $z=1$ of coherent and weak coupling with the environment, i.e. small dissipation constant $\gamma$, can help to achieve high success probabilities of both protocols. Motivated by the previous observation, now we plot Fig. \hyperref[fig:Fig03]{3(c)}, and we can notice that a small dissipation constant helps to achieve high success probabilities of both the protocols.

\section{A general method for selecting quantum channels for other variants of the proposed protocols\label{sec:general-method}}
There are several variants of all the existing quantum communication protocols. Specifically, several variants of remote state preparation (RSP) are well studied. For example, we may mention that controlled RSP, joint RSP, controlled joint RSP, bidirectional RSP, etc. (for more details see Ref. \cite{SSB+15} and references therein) are variants of the scheme for RSP. Likewise, there can also be several variants of the proposed protocols for remote implementation of hidden or partially unknown quantum operators. In other words, there may be variants of the protocols detailed in Section \ref{sec:hidden} and Section \ref{sec:unknown}. Different possible variants of the proposed protocol is discussed in Section \ref{sec:variants-riho-ripuo}. Also technique for optimizing controllers qubit has been discussed in Section \ref{sec:optimizing_controller_qubit}.

\subsection{Different possible variants of the RIHO and RIPUO protocols\label{sec:variants-riho-ripuo}}
It is to be noted the quantum channel used for both the proposed protocols is the same and is given by Eq. (\ref{eq:QC}). Interestingly, the same type of quantum channel can be used for executing different variants of the proposed protocols. In what follows, we will show that the proposed protocols can be generalized by modifying only a few steps of the original protocols.\\
We may consider that there are $M$-joint parties who jointly want to implement an operator on an arbitrary quantum state placed at a remote location. This may be viewed as the joint multiparty version of the RIHO or RIPUO. The general form of the quantum channel that can be used for the same can be given as follows:
\begin{equation}\label{eq:J-QC}
	|\Upsilon\rangle_{AB^1B^2...B^M}=(|a_0\rangle_A \otimes_{i=1}^{M}|b_0^{i}\rangle_{B^i} + |a_1\rangle_A \otimes_{i=1}^{M}|b_1^{i}\rangle_{B^i}).
\end{equation}

Similarly, if we consider that there are $N$ controllers along with the $M$ joint parties then the quantum channel to be used to complete the controlled-joint task is given as follows:
\begin{equation}\label{eq:JC-QC}
	|\Theta\rangle_{AB^1B^2...B^MC^1C^2...C^N}=(|a_0\rangle_A \otimes_{i=1}^{M}|b_0^{i}\rangle_{B^i}\otimes_{j=1}^{N}|c_0^{j}\rangle_{C^j} + |a_1\rangle_A \otimes_{i=1}^{M}|b_1^{i}\rangle_{B^i}\otimes_{j=1}^{N}|c_1^{j}\rangle_{C^j}).
\end{equation}
It is to be noted that the quantum channel given in Eq. (\ref{eq:JC-QC}) is the most general form for unidirectional RIHO and RIPUO protocol as the channels for other variants can be reduced from this. The task proposed can also be done in multi-directional ways. Like, multiple party can remotely implement operators on each others qubit simultaneously. Such scheme is called as multi-directional RIHO and RIPUO. Let there be $m$ such direction of communication then the quantum channel can be $|\Theta\rangle^{\otimes m}$. Protocols and strategies to be followed will be similar.\\
One can also think to reduce number of qubit used in the controlled version of the proposed protocol. Such technique to optimize controller qubit is discussed in the next section.

\subsection{Optimizing controller qubit of the controlled version of the RIHO and RIPUO protocol\label{sec:optimizing_controller_qubit}}
We would like to note that the protocols for controlled, cyclic controlled and bidirectional cyclic controlled versions of the tasks described above (i.e., RIHO and RIPUO) have been proposed so far \cite{A07,FL08,PH19,PY+22,PT+22}. Such a scheme requires multi-partite entangled states prepared by the controller as a quantum channel. After combining that with the state on which the remote operation is to be implemented we usually obtain a complex state of the form
 \begin{equation}
 	|\Psi\rangle =\frac{1}{\sqrt{2}}(|\chi_\mu\rangle|\mu\rangle+|\chi_\nu\rangle|\nu\rangle,
 \end{equation}
 where $\langle \chi_\mu|\chi_\nu\rangle = 0$ and $\langle \mu|\nu\rangle = 0$ (see Eq. (5) of Ref. \cite{PH19} and Eq. (9) of Ref. \cite{FL08,PY+22}  and the strategy used after that as an example \footnote{It is easy to generalize to multi-controller case. In that case, the combined state takes the form $|\psi\rangle =\sum_{\mu=0}^N\frac{1}{\sqrt{N}}(|\chi_\mu\rangle|\mu\rangle$, where $|\mu\rangle$ is viewed as a product state forming basis set $\{|\mu\rangle\}$ in $N$ dimension.}). At this stage, Charlie usually performs a measurement in $\{|\mu\rangle,|\nu\rangle\}$ basis (often the choice is computational basis, but in general it can be any basis) and discloses the outcome of the measurement. Knowledge of the outcome of Charlie's measurement allows Alice and Bob(s) to know the state they shared and to proceed accordingly. In the above case, if Charlie discloses that his measurement has revealed $|\mu\rangle  (|\nu\rangle)$, then Alice and Bob would know that they share $|\chi_\mu\rangle$  ($|\chi_\nu\rangle$) and will be able to execute the rest of the protocol accordingly. Now, there are two conceptual points to ponder at this moment. Firstly, in all such schemes for controlled computing or communication, Alice and Bob(s) are assumed to be semi-honest. Otherwise, they could have made states like $|\chi_\mu\rangle$  or $|\chi_\nu\rangle$ and execute the protocol themselves. Secondly, Charlie only discloses a bit of information to reveal the outcome of his measurement and thus to keep his control over the other semi-honest users. To do so in the standard approach he needs at least one qubit in his possession, and that qubit needs to be entangled with the states of Alice and Bob as only in that case his measurement will lead to different shared states and thus generate the power to controller. Such an approach leads to the requirement of large entangled state(s) which are difficult to create and maintain. Here, we note that Charlie can keep the control even without holding a qubit. Specifically, he may prepare the initial state randomly among a set of orthogonal states, say in a manner that the combined state of Alice and Bob becomes $|\chi_\mu\rangle$ or $|\chi_\nu\rangle$ depending on whether he started with a state $|\Psi_\mu\rangle=|\Omega^{+}\rangle^{\otimes n}$  or $|\Psi_\nu\rangle=|\Omega^{-}\rangle^{\otimes n}$  where $|\Omega^{\pm}\rangle=|a_0\rangle|b_0\rangle\pm|a_1\rangle|b_1\rangle$,  and $n$ is an integer. For the simplest version of the unidirectional protocols (i.e., for RIHO and RIPUO) and the versions with a single controller (i.e., for CRIHO and CRIPUO), the above state with $n=1$ would be sufficient. Now, in case of CRIHO and CRIPUO, instead of disclosing the measurement outcome, Charlie would disclose one bit of information stating whether he started with $|\Psi_\mu\rangle$ or $|\Psi_\nu\rangle$. The rest of the protocol would be the same and Charlie would have the same type of control over the semi-honest Alice and Bob(s) with the advantage of not requiring to prepare and maintain large entangled states. Similar results will be obtained if Charlie plans to start from  $|\Psi_\mu\rangle=|\Pi^{+}\rangle^{\otimes n}$ or $|\Psi_\nu\rangle=|\Pi^{-}\rangle^{\otimes n}$ where $|\Pi^{\pm}\rangle=|a_0\rangle|b_1\rangle\pm|a_1\rangle|b_0\rangle$. A closer look into the strategy described here shows that most of the related schemes can be realized using 1 or more Bell states which is relatively easy to prepare and maintain. For example, it will be sufficient to realize a scheme for bidirectional RIHO (BRIHO), bidirectional RIPUO (BRIPUO) and their controlled versions (i.e., CBRIHO and CBRIPUO) with two Bell states (i.e., by considering $n=2$ in the above-described states) and the above strategy, whereas for unidirectional (bidirectional) cyclic implementation of RIHO and RIPUO involving $m$ parties other than controller would require $m (2m)$ Bell states. This fact is summarized in Table \ref{tab:TableOptimize}, where we have listed the optimal resources to be required for realizing various RIO tasks using our ideas and have compared that with the quantum resources used for realizing similar tasks in the earlier works. The comparison clearly indicates the advantage of the present approach. Here, it would be apt to note that in the simplest version of the cyclic case, we will have $m=3$  and we may consider 3 users Alice, Bob and David. In this scenario,  Alice wants to remotely implement a hidden or partially unknown quantum operator on Bob's arbitrary quantum state and Bob and David want to perform similar tasks on David's and Alice's states, respectively.  The quantum channel that can be used to complete the task is three copies of the Bell states, say, $(|\Omega^{+}\rangle_{AB}|\Omega^{+}\rangle_{B'D}|\Omega^{+}\rangle_{D'A'})$. Qubits labeled as $A$ \& $A'$ is with Alice, $B$ \& $B'$ is with Bob and $D$ \& $D'$ is with David. Controlled version of this would require same amount of quanutm resource but bidirectional and controlled bidirectional versions of the cyclic RIPUO and cyclic RIHO would require two times more number of Bell states.
\begin{table}
\centering
\begin{tabular}{|c|c|c|c|c|}
\hline 
S.No & Quantum Channel Used & Purpose & Party & Optimal Channel\\
\hline 
1 & $\frac{1}{\sqrt{2}}(|000\rangle+|111\rangle)_{AA^{'}B}$ \cite{A07} & RIHO & 2 & $|\Omega^{\pm}\rangle$ or $|\Pi^{\pm}\rangle$\\
\hline
2 & $\frac{1}{\sqrt{2}}(|0000\rangle+|1111\rangle)_{AA^{'}BC}$ \cite{A07} & CRIHO & 2+1 & $|\Omega^{\pm}\rangle$ or $|\Pi^{\pm}\rangle$\\
\hline
3 & $\frac{1}{\sqrt{2}}(|000\rangle+|111\rangle)_{ABC}$ \cite{FL08} & CRIPUO & 2+1 & $|\Omega^{\pm}\rangle$ or $|\Pi^{\pm}\rangle$\\
\hline
4 & $|+\rangle|\phi^{+}\rangle^{\otimes3}+|-\rangle|\phi^{-}\rangle^{\otimes3}$ \cite{PH19} & CCRIPUO & 3+1 & $|\Omega^{\pm}\rangle^{\otimes3}$ or $|\Pi^{\pm}\rangle^{\otimes3}$\\
\hline
5 & $|0\rangle|\phi^{+}\rangle^{\otimes6}+|1\rangle|\phi^{-}\rangle^{\otimes6}$ \cite{PY+22} & BCCRIPUO & 3+1 & $|\Omega^{\pm}\rangle^{\otimes6}$ or $|\Pi^{\pm}\rangle^{\otimes6}$\\
\hline
\end{tabular}
\caption{Optimal resources for the proposed protocols for variants of RIHO and RIPUO. Here, $|\Omega^{\pm}\rangle=|a_0\rangle|b_0\rangle\pm|a_1\rangle|b_1\rangle$, $|\Pi^{\pm}\rangle=|a_0\rangle|b_1\rangle\pm|a_1\rangle|b_0\rangle$.}
\label{tab:TableOptimize}
\end{table}

Here, it is worth noting that, with regard to the above-mentioned strategy for optimizing the quantum resource, let us try to optimize the controllers' qubit given in Eq. (\ref{eq:JC-QC}) as it is the most general form of the quantum channel for the RIHO and RIPUO protocol. One of the $N$ controllers (say, controller $n$) may preserve his/her ability to control the working of the scheme to be implemented by Alice and Bob in a classical manner. Namely, that controller $n$ picks up a random bit $r_n\in\{0,1\}$, prepares an $M+N$ (not $M+N+1$ as in Eq. (\ref{eq:JC-QC})) partite entangled state       
\begin{equation}\label{eq:JC-QC1}
    |\Theta^{'}\rangle_{AB^1B^2...B^MC^1...C^{n-1}...C^{n+1}...C^N}=(|a_0\rangle_A \otimes_{i=1}^{M}|b_0^{i}\rangle_{B^i}\otimes_{j=1, j\neq n}^{N}|c_0^{j}\rangle_{C^j} + (-1)^{r_n} |a_1\rangle_A \otimes_{i=1}^{M}|b_1^{i}\rangle_{B^i}\otimes_{j=1, j\neq n}^{N}|c_1^{j}\rangle_{C^j}).
\end{equation}

and then lets the other participants share the thus prepared state. The precise value of $r_n$ will be broadcasted by the controller $n$ only at the end of the protocol. Qualitatively,  the capability of the controller remains unchanged, but the quantum resource saves one qubit and the cost for the controller's quantum measurement. Yet, the most optimal strategy could be achieved if the $N$ controllers cooperate in the following way. One of the controllers, say, controller $1$, prepares an $M + 2$ partite entangled state of the form

\begin{equation}\label{eq:JC-QC2}
    |\Theta^{''}\rangle_{AB^1B^2...B^MC}=(|a_0\rangle_A \otimes_{i=1}^{M}|b_0^{i}\rangle_{B^i}|c_0\rangle_{C} + (-1)^{r_1} |a_1\rangle_A \otimes_{i=1}^{M}|b_1^{i}\rangle_{B^i}|c_1\rangle_{C}).
\end{equation}
with a random bit $r_1\in\{0,1\}$. Then the controller $1$ lets Alice, $M$ Bobs and the controller $2$ share the state (Eq. \ref{eq:JC-QC2}) with qubit $C$ to be held by the controller $2$. The controller $2$ chooses either to apply or not the phase-flip operator on qubit $C$ (mathematically, this implies that the controller $2$ applies on qubit $C$ the operator $Z^{r_2}$ with $r_2$ a random bit, and then forwards qubit $C$ to the controller $3$ who will do the similar action like the controller $2$ did with a new random bit $r_3$. Next, the controller $3$ forwards qubit $C$ to the controller $4$ and so on until the controller $N$. As a consequence, the total state shared among the controller $1$, Alice and $M$ Bob becomes
\begin{equation}\label{eq:JC-QC3}
    |\Theta^{'''}\rangle_{AB^1B^2...B^MC}=(|a_0\rangle_A \otimes_{i=1}^{M}|b_0^{i}\rangle_{B^i}|c_0\rangle_{C} + (-1)^{r} |a_1\rangle_A \otimes_{i=1}^{M}|b_1^{i}\rangle_{B^i}|c_1\rangle_{C}).
\end{equation}
with $r=\sum_{j=1}^{N}r_j$. During the protocol execution the controller $1$ needs to measure qubit $C$ with the outcome $r_0\in\{0,1\}$. At an appropriate step near the end of the protocol the controller $1$ discloses both $r_0$ and $r_1$, while each of the remaining controllers discloses his/her random bit $r_j$. This allows Alice and $M$ Bobs know the exact values of $r_0$ and $r$ to properly complete the protocol. Each controller maintains his/her independent power and the cost for the quantum resource is reduced considerably.
\section{Summary and Conclusion\label{sec:conclusion}}

We have shown that a family of protocols for different variants of RIO can be realized using optimal resources. Among these protocols, two protocols are discussed elaborately. To be precise, a protocol for RIHO and another protocol for RIPUO using optimal resources are described in detail as well the dissipation of auxiliary coherent state used for cross-Kerr nonlinear interaction has been studied. Also, it is shown that these protocols can be easily generalized to their joint, controlled, controlled joint, bidirectional, cyclic, controlled cyclic, and controlled bidirectional versions. We have also shown that without giving any extra qubit(s) to the controller, one can realize the controlled versions of the RIHO and RIPUO using the same amount of quantum resources as needed for the corresponding protocol without the controller.  Any protocol that implements a variant of RIO in a controlled manner essentially demands that Alice and Bob are semi-honest and possess some ignorance about the state they share (usually 1 bit or more). This ignorance can be removed by the controller (Charlie) by disclosing an equal amount of classical information (usually by performing a measurement and disclosing the outcome). Here, we have shown that exploiting the fact that RIHO and RIPUO can be implemented using different Bell states with minor changes in the steps (as summarized in Tables \ref{tab:stepsRIHO} and \ref{tab:stepsRIPUO}), controlled versions of RIHO and RIPUO can be prepared where Charlie keeps Alice and Bob ignorant about the Bell state they share till a particular step after which the steps (actions to be taken by them) will be different for different initial states prepared by Charlie. This strategy allows us to realize several schemes with optimal resources as summarized in Table \ref{tab:TableOptimize}, which clearly indicates that the large entangled states used in previous works were not essential for the execution of the corresponding tasks. As Bell states of the forms used here are experimentally realizable and relatively easier to maintain, we are hopeful that the proposed schemes will be experimentally realized soon. The expectation related to experimental realizability is enhanced with the fact that in the present work, we have shown that the operators that we implement remotely can be viewed as single qubit rotation and in Ref. \cite{XLG05} an experimental realization of remote implementation of rotation operator is already reported.

Finally, we would like to conclude this paper by noting that there are many ways, in which present work can be extended in the near future. Specifically, extending the works of Ref. \cite{ShohiniG2015ControlPower,BaAn2019ControlPower} control power can be computed. Further, till now what we have discussed is the remote implementation of an operator between two nodes of a network that share an entangled state. What happens if Alice and Bob are part of a network, but they do not share an entangled state? Such a question is discussed in Ref. \cite{WTB+22} in the context of quantum multihop networks. However, without restricting us to quantum multihop networks and simple RIO we can perform different variants of RIO in more general networks as our schemes require only Bell states and efficient schemes for entanglement routing over more general networks have recently been proposed by some of the present authors \cite{MP23,MP22}. Keeping these possibilities in mind, we conclude this work with a hope that our proposed protocols will be experimentally realized and lead to many applications in more complex situations in the near future.
\section*{Acknowledgment}
Authors acknowledge the support from the Chanakya Doctoral Fellowship program of I-HUB Quantum Technology Foundation (QTF), IISER Pune, India (Grant No.: I-HUB/QUEST-DF/2023-24/09). Authors are also thankful to Nguyen Ba An for his interest in this work and the technical suggestions provided by him.

\section*{Data Availability} Data sharing is not applicable to this article as no datasets were generated or analyzed during the current study.

\bibliographystyle{unsrt}
\bibliography{riho}

\end{document}